\documentclass[a4paper]{jpconf}
\usepackage{graphicx}
\usepackage{clshan-math,clshan-dm-dd}
\newcommand{\plotsinsert} [5]
{\begin{figure}[#1]
  \begin{center}
   \hspace*{-1.6cm}
   \includegraphics[width=8cm]{#2-bg-#3} \hspace{0.25cm}
   \includegraphics[width=8cm]{#2-bg-#4} \hspace*{-1.6cm} \\
   \vspace{-0.25cm}
  \end{center}
  \caption{#5}
  \label{fig:#2}
 \end{figure}
}
\def \lsim {\:\raisebox{-0.7ex}{$\stackrel{\textstyle<}{\sim}$}\:}

\begin{document}
\title{Analyzing direct dark matter detection data \\ \vspace{-0.15cm}
       with unrejected background events           \\ \vspace{-0.15cm}
       by the AMIDAS website}
\author{Chung-Lin Shan}
\address{\it Institute of Physics, Academia Sinica    \\
             No.~128, Sec.~2, Academia Road, Nankang,
             Taipei 11529, Taiwan, R.O.C.}
\ead{clshan@phys.sinica.edu.tw}
\begin{abstract}
 In this talk
 I have presented the data analysis results
 of extracting properties of halo WIMPs:
 the mass and the (ratios between the) spin--independent
 and spin--dependent couplings/cross sections on nucleons
 by the \amidas\ website
 by taking into account
 possible {\em unrejected} background events in the analyzed data sets.
 Although {\em non--standard} astronomical setup
 has been used
 to generate {\em pseudodata} sets
 for our analyses,
 it has been found that,
 {\em without} prior information/assumption
 about the local density and velocity distribution of halo Dark Matter,
 these WIMP properties have been reconstructed with
 $\sim$ 2\% to $\lsim$ 30\% deviations from the input values.
\end{abstract}
\section{Introduction}
 In order to extract properties of halo WIMPs
 (Weakly Interacting Massive Particles)
 by using data from direct Dark Matter detection experiments
 as model--independently as possible,
 we have developed a series of data analysis method
 for reconstructing the one--dimensional
 WIMP velocity distribution
 \cite{DMDDf1v}
 as well as determining the WIMP mass
 \cite{DMDDmchi},
 the spin--independent (SI) WIMP coupling on nucleons
 \cite{DMDDfp2}
 and
 the ratios between different WIMP couplings/cross sections
 \cite{DMDDranap}.
 Moreover,
 in collaboration with
 the DAMNED (DArk Matter Network Exclusion Diagram) Dark Matter online tool
 \cite{DAMNED},
 part of the ILIAS Project \cite{ILIAS},
 the ``\amidas'' (A Model--Independent Data Analysis System) website
 for online simulation/data analysis
 has also been established
 \cite{AMIDAS-web, AMIDAS-SUSY09, AMIDAS-f1vFQ}.

 In this article,
 in order to demonstrate the usefulness and powerfulness
 as well as the model--independence of the \amidas\ package
 for direct Dark Matter detection experiments,
 I will analyze {\em blindly} some {\em pseudodata} sets
 generated theoretically for different detector materials
 and present the reconstructed WIMP properties.
 This means that
 I will simply upload these data sets onto the \amidas\ website
 and follow the instructions to reconstruct different WIMP properties
 {\em without} using any information about the input setup
 used for generating the analyzed pseudodata.
 For cases in which
 some information about WIMPs (e.g., the mass $\mchi$)
 and/or Galactic halo (e.g., the local Dark Matter density $\rho_0$)
 is required,
 I will naively use the commonly used/favorite values
 for the data analyses.
 Moreover,
 due to the time limit on preparing
 the presentation for the CYNGUS 2011 workshop,
 in Ref.~\cite{AMIDAS-CYGNUS2011}
 we considered only the case of {\em null} background events.
 In this article,
 we have taken into account
 a small fraction of
 possible {\em unrejected} background events in the analyzed data sets
 \cite{DMDDbg-mchi, DMDDbg-f1v, DMDDbg-fp2, DMDDbg-ranap}.

 After that
 I show the blindly reconstructed properties of halo WIMPs
 in Sec.~2,
 in Sec.~3
 I will reveal the input setup and the background spectrum
 used for generating the analyzed data
 and compare the reconstructed results to them.
 Finally,
 I conclude in Sec.~4.
\section{Reconstructed WIMP properties}
 In this section,
 I present the reconstructed WIMP properties
 analyzed by the \amidas\ website.
 While in each uploaded file
 there are {\em exactly} 50 data sets,
 in each data set
 there are {\em on average} 50 recorded events
 (i.e., 50 measured recoil energies);
 the exact number of total events is Poisson distributed.
 For simplicity,
 the experimental minimal and maximal cut--off energies
 have been set as 0 and 100 keV
 for all data sets.

 In order to check the effect of
 using a ``wrong'' elastic nuclear form factor,
 two forms have been considered
 for the SI WIMP--nucleus cross section
 in our analyses.
 One is the simple exponential form:
\beq
   F_{\rm ex}^2(Q)
 = e^{-Q / Q_0}
\~.
\label{eqn:FQ_ex}
\eeq
 Here $Q$ is the recoil energy
 transferred from the incident WIMP to the target nucleus,
 $Q_0$ is the nuclear coherence energy given by
\(
   Q_0
 = 1.5 / \mN R_0^2
\),
 where
\(
   R_0
 = \bbrac{0.3 + 0.91 \abrac{\mN / {\rm GeV}}^{1/3}} \~ {\rm fm}
\)
 is the radius of the nucleus
 and $\mN$ is the mass of the target nucleus.
 Meanwhile,
 we used also a more realistic analytic form
 for the elastic nuclear form factor:
\beq
   F_{\rm SI}^2(Q)
 = \bfrac{3 j_1(q R_1)}{q R_1}^2 e^{-(q s)^2}
\~.
\label{eqn:FQ_WS}
\eeq
 Here $j_1(x)$ is a spherical Bessel function,
\(
   q
 = \sqrt{2 m_{\rm N} Q}
\)
 is the transferred 3-momentum,
 for the effective nuclear radius we use
\beq
   R_1
 = \sqrt{R_A^2 - 5 s^2}
\label{eqn:R1}
\eeq
 with
\(
        R_A
 \simeq 1.2 \~ A^{1/3}~{\rm fm}
\)
 and a nuclear skin thickness
\(
        s
 \simeq 1~{\rm fm}
\).
 For the SD WIMP--nucleus cross section,
 we only used the ``thin--shell'' nuclear form factor:
\beqn
    F_{\rm TS}^2(Q)
 \= \cleft{\renewcommand{\arraystretch}{1.15}
           \begin{array}{l l l}
            j_0^2(q R_1)                      \~, & ~~~~~~~~ &
            {\rm for}~q R_1 \le 2.55~{\rm or}~q R_1 \ge 4.5 \~, \\ 
            {\rm const.} \simeq 0.047         \~, &          &
            {\rm for}~2.5 5 \le q R_1 \le 4.5 \~.
           \end{array}}
\label{eqn:FQ_TS}
\eeqn
\subsection{WIMP mass $\mchi$}
\plotsinsert{t!}{mchi}{ex}{WS}
{The reconstructed WIMP mass $\mchi$
 by means of the method introduced
 in Ref.~\cite{DMDDmchi}
 with a target combination of $\rmXA{Si}{28}$ + $\rmXA{Ge}{76}$ nuclei.
 Two forms of the elastic nuclear form factor
 given in Eqs.~(\ref{eqn:FQ_ex}) and (\ref{eqn:FQ_WS})
 have been used
 in the left and right frames,
 respectively.
}

 As one of the most important properties of halo WIMPs
 as well as the basic information for reconstructing other quantities
 in our model--independent analysis methods,
 I consider at first
 the determination of the WIMP mass $\mchi$
 by means of the method introduced
 in Ref.~\cite{DMDDmchi}.

 In Figs.~\ref{fig:mchi}
 I show the reconstructed WIMP masses and
 the lower and upper bounds of their 1$\sigma$ statistical uncertainties.
 The usual target combination of
 $\rmXA{Si}{28}$ + $\rmXA{Ge}{76}$ nuclei
 has been used for this reconstruction,
 whereas
 two analytic forms of the elastic nuclear form factor
 given in Eqs.~(\ref{eqn:FQ_ex}) and (\ref{eqn:FQ_WS})
 have been used for determining $\mchi$
 in the left and right frames,
 respectively.
 While $m_{\chi, n}$ with $n = -1$, 1, 2 and $m_{\chi, \sigma}$
 have been estimated by Eqs.~(34) and (40) of Ref.~\cite{DMDDmchi},
 respectively,
 $m_{\chi, {\rm combined}}$ has been estimated
 by the $\chi^2$--fitting
 defined in Eq.~(51) of Ref.~\cite{DMDDmchi},
 which combines the estimators for
 $m_{\chi, n}$ and $m_{\chi, \sigma}$ with each other.
 The reconstructed WIMP mass $m_{\chi, {\rm combined}}$
 as well as $m_{\chi, n}$ and $m_{\chi, \sigma}$
 shown here have been corrected
 by the iterative $\Qmax$--matching procedure
 described in Ref.~\cite{DMDDmchi}.

 It can be found here that,
 although all single estimators
 ($m_{\chi, n}$ with $n = -1$, 1, 2 and $m_{\chi, \sigma}$)
 give generally a (relatively lighter) WIMP mass of
 \mbox{$\sim 60$ GeV} or even lighter
 and a 1$\sigma$ upper bound of $\sim 140$ GeV,
 the {\em mean} values of the combined
 (in principle, more reliable) results
 (second columns)
 of the reconstructed WIMP mass give $\mchi \sim 115$ GeV
 with a {\em rough} 1$\sigma$ upper (lower) bound of $\sim 190$ (80) GeV,
 or, equivalently,
\beq
        \mchi
 \simeq 115_{-35}^{+75}~{\rm GeV}
\~.
\label{eqn:mchi_rec}
\eeq
 Moreover,
 the combined results
 with two different form factors
 show not only a large overlap between $\sim 85$ GeV and $\sim 180$ GeV,
 but also a good coincidence:
 comparing to the $\sim_{-35}^{+75}$ GeV
 1$\sigma$ statistical uncertainty
 and the $\sim_{-30}^{+45}$ GeV overlap,
 the difference between two median values is $\lsim~30$ GeV!
 This indicates that,
 for the first approximation of
 giving/constraining the most plausible range of the WIMP mass
 with pretty few total events,
 the uncertainty on the nuclear form factor
 could be safely neglected.

\subsection{Spin--independent WIMP--nucleon coupling $|f_{\rm p}|^2$}
\plotsinsert{t!}{fp2}{03-ex}{input-WS}
{The reconstructed {\em squared} SI WIMP--nucleon coupling $|f_{\rm p}|^2$
 by means of the method introduced
 in Ref.~\cite{DMDDfp2}
 with a $\rmXA{Ge}{76}$ target.
 The commonly used value of
 the local Dark Matter density
 $\rho_0 = 0.3~{\rm GeV/cm^3}$
 and a larger value of $\rho_0 = 0.4~{\rm GeV/cm^3}$
 as well as
 the elastic nuclear form factors
 given in Eqs.~(\ref{eqn:FQ_ex}) and (\ref{eqn:FQ_WS})
 have been used for estimating $|f_{\rm p}|^2$
 in the left and right frames,
 respectively.
}

 Following the WIMP mass determination,
 I consider now the reconstruction of
 the SI WIMP coupling on nucleons $|f_{\rm p}|^2$
 with a $\rmXA{Ge}{76}$ target
 \cite{DMDDfp2}
\footnote{
 Remind that
 the theoretical prediction
 by most supersymmetric models that
 the SI scaler WIMP couplings
 on protons and on neutrons are (approximately) equal:
 $f_{\rm p} \simeq f_{\rm n}$
 has been adopted in the \amidas\ package.
}.

 In Figs.~\ref{fig:fp2}
 I show the reconstructed squared SI WIMP-nucleon couplings
 and the lower and upper bounds of
 their 1$\sigma$ statistical uncertainties
 estimated by Eqs.~(17) and (18) of Ref.~\cite{DMDDfp2}
 with an {\em assumed} (100$\pm$10 GeV, labeled with the subscript ``input'')
 and the reconstructed (from Sec.~2.1, labeled with ``recon'') WIMP masses.
 The commonly used value of
 the local Dark Matter density
 $\rho_0 = 0.3~{\rm GeV/cm^3}$
 and a larger value of $\rho_0 = 0.4~{\rm GeV/cm^3}$
 \cite{Catena09, Salucci10, Pato10}
 as well as
 the elastic nuclear form factors
 given in Eqs.~(\ref{eqn:FQ_ex}) and (\ref{eqn:FQ_WS})
 have been used for estimating $|f_{\rm p}|^2$
 in the left and right frames,
 respectively.

 Among these results,
 the {\em mean} value and the {\em overlap} of two most plausible results
 (estimated with the reconstructed WIMP mass)
 give a {\em rough} 1$\sigma$ range of
\beq
        |f_{\rm p}|^2
 \simeq 1.05_{-0.24}^{+0.33} \times 10^{-17}~{\rm GeV}^{-4}
\~,
\label{eqn:fp2_rec}
\eeq
 or, equivalently,
\beq
        |f_{\rm p}|
 \simeq 3.24_{-0.37}^{+0.51} \times 10^{-9}~{\rm GeV}^{-2}
\~.
\label{eqn:fp_rec}
\eeq
 Since the reconstructed WIMP mass given in Sec.~2.1 is
 $\mchi \sim 115$ GeV,
 one can simply use the proton mass $m_{\rm p}$
 to approximate the WIMP--proton reduced mass $\mrp$
 and give a reconstructed SI WIMP--nucleon cross section as
 \cite{AMIDAS-CYGNUS2011}
\beq
         \sigmapSI
 =       \afrac{4}{\pi} \mrp^2 |f_{\rm p}|^2
 \approx \afrac{4}{\pi} m_{\rm p}^2 \~ |f_{\rm p}|^2
 \simeq  5.03_{-1.15}^{+1.58} \times 10^{-9}~{\rm pb}
\~.
\label{eqn:sigmapSI_rec}
\eeq
\subsection{Ratio of two spin--dependent WIMP--nucleon couplings
            $\armn / \armp$}
\plotsinsert{t!}{ranap}{ex}{WS}
{The reconstructed ratio between two SD WIMP--nucleon couplings,
 $\armn / \armp$,
 by means of two methods introduced
 in Ref.~\cite{DMDDranap}.
 As usual,
 the elastic nuclear form factors
 given in Eqs.~(\ref{eqn:FQ_ex}) and (\ref{eqn:FQ_WS})
 have been used for determining $\armn / \armp$
 in the left and right frames,
 respectively.
}

 In Figs.~\ref{fig:ranap}
 I show the reconstructed $\armn / \armp$ ratios
 and the lower and upper bounds of
 their 1$\sigma$ statistical uncertainties
 estimated by Eqs.~(2.7) and (2.12) of Ref.~\cite{DMDDranap} with $n = 1$
 as well as by Eqs.~(3.16) and (3.20) of Ref.~\cite{DMDDranap}
 at the shifted energy points
 \cite{DMDDf1v, DMDDranap}.
 A combination of \mbox{$\rmXA{F}{19}$ + $\rmXA{I}{127}$} targets
 has been used
 for the reconstruction of $\armn / \armp$
 under the assumption that
 the SD WIMP--nucleus interaction dominates over the SI one
 (labeled with the superscript ``SD''),
 whereas
 a third target of $\rmXA{Si}{28}$ has been combined
 with $\rmXA{F}{19}$ and $\rmXA{I}{127}$
 for the case of the general combination of both SI and SD WIMP interactions
 (labeled with the superscript ``SI + SD'').

 It can be found that,
 firstly,
 the ``$+$ (plus)'' solutions of the $\armn / \armp$ ratios given here
 are obviously too large to be the reasonable choice for $\armn / \armp$
 and the ``$-$ (minus)'' solutions should be the correct ones%
\footnote{
 Remind that,
 as discussed in Ref.~\cite{DMDDranap},
 the correct choice from the ``$+$'' and ``$-$'' solutions
 could be decided directly by the values of
 the group spins of protons and neutrons of the used target nuclei,
 $\expv{S_{\rm (p, n)}}$.
}.
 Secondly,
 although the reconstructed results
 under the assumption of the SD dominant WIMP interaction
 (third columns)
 is in general {\em larger} than
 the (in principle more plausible) results
 obtained without such a prior assumption (last columns),
%
%
 one could still use
 the {\em mean} value and the {\em overlap} of these two results
 to give a {\em rough} 1$\sigma$ range of
\beq
        \frac{\armn}{\armp}
 \simeq 0.65_{-0.30}^{+0.23}
\~.
\label{eqn:ranap_rec}
\eeq
\subsection{Ratios of the SD and SI WIMP--nucleon couplings
            $\sigma_{\chi ({\rm p, n})}^{\rm SD} / \sigmapSI$}
\plotsinsert{t!}{rsigmaSDSI}{ex}{WS}
{The reconstructed ratios between the SD and SI WIMP--nucleon couplings,
 $\sigma_{\chi ({\rm p, n})}^{\rm SD} / \sigmapSI$,
 by means of two methods introduced
 in Ref.~\cite{DMDDranap}.
 As usual,
 the elastic nuclear form factors
 given in Eqs.~(\ref{eqn:FQ_ex}) and (\ref{eqn:FQ_WS})
 have been used for determining
 $\sigma_{\chi ({\rm p, n})}^{\rm SD} / \sigmapSI$
 in the left and right frames,
 respectively.
}

 In Figs.~\ref{fig:rsigmaSDSI}
 I show the reconstructed
 $\sigma_{\chi ({\rm p, n})}^{\rm SD} / \sigmapSI$ ratios
 and the lower and upper bounds of
 their 1$\sigma$ statistical uncertainties
 estimated by Eqs.~(3.9), (3.10) and (3.21) of Ref.~\cite{DMDDranap}
 (with $\armn / \armp$ estimated by Eq.~(3.16) of Ref.~\cite{DMDDranap})
 as well as by Eqs.~(3.25) and (3.29) of Ref.~\cite{DMDDranap}
 at the shifted energy points.
 A combination of data sets of
 $\rmXA{F}{19}$, $\rmXA{I}{127}$ and $\rmXA{Si}{28}$ targets
 (labeled with the superscript ``XYZ'')
 and that of data sets of $\rmXA{Na}{23}$ or $\rmXA{Xe}{131}$
 with the (common) one of $\rmXA{Ge}{76}$
 (labeled with the superscript ``XY'')
 have been used and
 the {\em mean} value and the {\em overlap} of these two results
 give a {\em rough} 1$\sigma$ range of
\beq
        \frac{\sigmapSD}{\sigmapSI}
 \simeq 8.94_{-2.67}^{+2.13} \times 10^5
\~,
        ~~~~~~~~~~~~~~~~ 
        \frac{\sigmanSD}{\sigmapSI}
 \simeq 3.16_{-1.05}^{+2.36} \times 10^5
\~.
\label{eqn:rsigmaSDSI_rec}
\eeq
 Then,
 firstly,
 from these results one can further obtain that
 \cite{AMIDAS-CYGNUS2011}%
\footnote{
 Remind that
 the results given in the second and third columns
 of the tables in Figs.~\ref{fig:rsigmaSDSI}
 are reconstructed with the $\armn / \armp$ ratio
 given in the last columns of the tables in Figs.~\ref{fig:ranap}.
}
\beq
        \vfrac{\armn}{\armp}
 \simeq 0.59_{-0.13}^{+0.23}
\~.
\label{eqn:ranap_rec_2}
\eeq
 Secondly,
 combining the results
 in Eq.~(\ref{eqn:rsigmaSDSI_rec})
 with $\sigmapSI$ given in Eq.~(\ref{eqn:sigmapSI_rec}),
 one can obtain that \cite{AMIDAS-CYGNUS2011}
\beq
        \sigmapSD
 \simeq 4.50_{-1.69}^{+1.77} \times 10^{-3}~{\rm pb}
\~,
        ~~~~~~~~~~~~~~~~ 
        \sigmanSD
 \simeq 1.59_{-0.64}^{+1.08} \times 10^{-3}~{\rm pb}
\~.
\label{eqn:rsigmaSD_rec}
\eeq
 These results give in turn that \cite{AMIDAS-CYGNUS2011}
\beq
   |\armp|
 = 0.112_{-0.021}^{+0.022}
\~,
   ~~~~~~~~~~~~~~~~ 
   |\armn|
 = 0.067_{-0.013}^{+0.023}
\~.
\label{eqn:a_rec}
\eeq
 On the other hand,
 one can also use the reconstructed $\armn / \armp$ ratio
 given in Eq.~(\ref{eqn:ranap_rec})
 and {\em one} of the two results
 given in Eq.~(\ref{eqn:rsigmaSD_rec})
 to obtain that \cite{AMIDAS-CYGNUS2011}
\beq
        \sigmapSD
 \simeq 3.76_{-3.81}^{+3.69} \times 10^{-3}~{\rm pb}
\~,
        ~~~~~~~~~~~~~~~~ 
        \sigmanSD
 \simeq 1.90_{-1.89}^{+1.54} \times 10^{-3}~{\rm pb}
\~.
\label{eqn:rsigmaSD_rec_2}
\eeq
 These results can also give that
\beq
        |\armp|
 \simeq 0.103_{-0.052}^{+0.050}
\~,
        ~~~~~~~~~~~~~~~~ 
        |\armn|
 \simeq 0.073_{-0.036}^{+0.030}
\~.
\label{eqn:a_rec_2}
\eeq

 It can be found that,
 not surprisingly,
 the statistical uncertainties
 on the reconstructed $\sigma_{\chi ({\rm p, n})}^{\rm SD}$
 given in Eq.~(\ref{eqn:rsigmaSD_rec_2})
 are $\sim$ 2 or 3 times larger than those
 given in Eq.~(\ref{eqn:rsigmaSD_rec}):
 Since $\sigma_{\chi ({\rm p, n})}^{\rm SD} / \sigmapSI$
 reconstructed with the F + I + Si combination
 involve already the reconstructed $\armn / \armp$ ratio
 given in Eq.~(\ref{eqn:ranap_rec}),
 the uncertainties on $\sigma_{\chi ({\rm p, n})}^{\rm SD}$
 given in Eq.~(\ref{eqn:rsigmaSD_rec_2})
 are thus {\em overestimated}.
 Secondly,
 the reconstructed $\armn / \armp$ ratio
 given in Eqs.~(\ref{eqn:ranap_rec}) and (\ref{eqn:ranap_rec_2})
 and the reconstructed $\sigma_{\chi ({\rm p, n})}^{\rm SD}$
 given in Eqs.~(\ref{eqn:rsigmaSD_rec}) and (\ref{eqn:rsigmaSD_rec_2})
 seem to match to each other pretty well.

 The analyses given here show that,
 firstly,
 once one can estimate
 the SI WIMP--nucleon coupling/cross section,
 $|f_{\rm p}|$ or $\sigmapSI$,
 and (one of) the ratios between
 the SD and SI WIMP--nucleon cross sections,
 and/or the ratio between two SD WIMP--nucleon couplings,
 the other couplings/cross sections
 could in principle be estimated.
 Secondly,
 the WIMP couplings/cross sections estimated in different way
 would be self--cross--checks to each other and
 the (in)compatibility between the reconstructed results
 would also help us to check
 the usefulness of the analyzed data sets
 offered from different experiments with different detector materials
 \cite{DMDDbg-ranap}.

\section{Input setup for generating pseudodata}

 In Table \ref{tab:setup}
 I give finally the input setup
 for generating the pseudodata sets used in the analyses
 demonstrated in the previous section.
 For comparison,
 the reconstructed results shown in the previous section
 are also summarized here.

 For generating WIMP signals,
 we used the commonly used form for the nuclear form factor
 given in Eq.~(\ref{eqn:FQ_WS}) with
 another often used analytic form for $R_1$:
\beq
   R_1
 = \sqrt{R_A^2 + {\T \afrac{7}{3}} \pi^2 r_0^2 - 5 s^2}
\label{eqn:R1_Helm}
\eeq
 with
\(
        R_A
 \simeq \abig{1.23 \~ A^{1/3} - 0.6}~{\rm fm}
\),
\(
        r_0
 \simeq 0.52~{\rm fm}
\),
\(
        s
 \simeq 0.9~{\rm fm}
\).
 Moreover,
 the shifted Maxwellian velocity distribution:
\beq
   f_{1, \sh}(v)
 = \frac{1}{\sqrt{\pi}} \afrac{v}{\ve v_0}
   \bbigg{ e^{-(v - \ve)^2 / v_0^2} - e^{-(v + \ve)^2 / v_0^2} }
\label{eqn:f1v_sh}
\eeq
 with the Sun's Galactic orbital velocity $v_0 = 230$ km/s
 has been used;
 $\ve$ is the {\em time--dependent} Earth's velocity
 in the Galactic frame:
\beq
   \ve(t)
 = v_0 \bbrac{1.05 + 0.07 \cos\afrac{2 \pi (t - t_{\rm p})}{1~{\rm yr}}}
\~,
\label{eqn:ve}
\eeq
 where the date
 on which the Earth's velocity relative to the WIMP halo
 is maximal
 has been set as $t_{\rm p} = 140$ d.
 In addition,
 different from our setup used in Ref.~\cite{AMIDAS-CYGNUS2011},
 the experimental running date has been set as
 $t_{\rm expt} = 100$ d
 and thus $\ve(t_{\rm expt}) = 254$ km/s,
 much larger than the usually used values:
 $200~{\rm km/s} \le \ve \le 240~{\rm km/s}$.
 Although these values for the astronomical setup
 are {\em non--standard},
 we would like to stress that,
 as shown in the previous section and \mbox{Table I},
 such a non--standard halo (model)
 would not affect the reconstructed results,
 since
 for using the \amidas\ package and website
 to analyze (real) data sets,
 one needs only the form factors
 for SI and/or SD WIMP--nucleaus cross sections,
 prior knowledge/assumptions about
 the WIMP velocity distribution $f_1(v)$
 and local density $\rho_0$
 (except the estimation of the SI WIMP--nucleon coupling $|f_{\rm p}|^2$)
 are {\em not required}.

\begin{table}[t!]
\begin{center}
\renewcommand{\arraystretch}{0.3}
\begin{tabular}{|| c | c | c | l ||}
\hline
\hline
 & & & \\
 \makebox[2  cm][c]{Property}              &
 \makebox[4.5cm][c]{Reconstructed   value} &
 \makebox[4.5cm][c]{Input/Estimated value} &
 \makebox[1.5cm][c]{Remarks}               \\
 & & & \\
\hline
\hline
 & & & \\
 $\mchi$                   &
 $115_{-35}^{+75}$ GeV     &
  130              GeV     & \\
 & & & \\
\hline
\hline
 & & & \\
 $\sigmapSI$               &
 $5.03_{-1.15}^{+1.58}     \times 10^{-9}$ pb &
 $4                        \times 10^{-9}$ pb &
 $f_{\rm n} = f_{\rm p}$   \\
 & & & \\
\hline
 & & & \\
 $|f_{\rm p}|^2$           &
 $1.05_{-0.24}^{+0.33}     \times 10^{-17}~{\rm GeV}^{-4}$ &
 $9.305                    \times 10^{-18}~{\rm GeV^{-4}}$ &
 $\dagger$                 \\
 & & & \\
\hline
 & & & \\
 $|f_{\rm p}|$             &
 $3.24_{-0.37}^{+0.51}     \times 10^{-9 }~{\rm GeV}^{-2}$ &
 $3.050                    \times 10^{-9 }~{\rm GeV}^{-2}$ &
 $\dagger$                 \\
 & & & \\
\hline
\hline
 & & & \\
 $\armp$                   &
 $0.112_{-0.021}^{+0.022}$ &
  0.1                      & \\
 & & & \\

 $\armn$                   &
 $0.067_{-0.013}^{+0.023}$ &
  0.07                     & \\
 & & & \\
\hline
 & & & \\
 $\armn / \armp$           &
 $0.65_{-0.30}^{+0.23}$,
 $0.59_{-0.13}^{+0.23}$    &
  0.7                      & \\
 & & & \\
\hline
 & & & \\
 $\sigmapSD$               &
 $4.50_{-1.69}^{+1.77}     \times 10^{-3}$ pb &
 $3.51                     \times 10^{-3}$ pb &
 $\dagger$                 \\
 & & & \\
 & & & \\
 $\sigmanSD$               &
 $1.59_{-0.64}^{+1.08}     \times 10^{-3}$ pb &
 $1.72                     \times 10^{-3}$ pb &
 $\dagger$                 \\
 & & & \\
\hline
 & & & \\
 $\sigmapSD / \sigmapSI$   &
 $8.94_{-2.67}^{+2.13}     \times 10^5$ &
 $8.77                     \times 10^5$ &
 $\dagger$                 \\
 & & & \\
 & & & \\
 $\sigmanSD / \sigmapSI$   &
 $3.16_{-1.05}^{+2.36}     \times 10^5$ &
 $4.30                     \times 10^5$ &
 $\dagger$                 \\
 & & & \\
\hline
\hline
 & & & \\
 $\FSIQ$                   & & $\FSIQ$ in Eq.~(\ref{eqn:FQ_WS})           & $\ddagger$ \\
 & & & \\
 $\FSDQ$                   & & $F_{\rm TS}^2(Q)$ in Eq.~(\ref{eqn:FQ_TS}) & \\
 & & & \\
\hline
\hline
 & & & \\
 $\rho_0$            & & $0.4~{\rm GeV/cm^3}$ & \\
 & & & \\
\hline
 & & & \\
 $t_{\rm p}$         & & 140 d                & \\
 & & & \\
 $t_{\rm expt}$      & & 100 d                & $\ddagger$ \\
 & & & \\
\hline
 & & & \\
 $v_0$               & & $230~{\rm km/s}$     & \\
 & & & \\
 $\vmax$             & & $600~{\rm km/s}$     & \\
 & & & \\
 $\ve(t_{\rm expt})$ & & $253.9~{\rm km/s}$   & $\ddagger$ \\
 & & & \\
\hline
\hline
 & & & \\
 $r_{\rm bg}$        & & 0.12                 & $\ddagger$ \\
 & & & \\
\hline
\hline
\end{tabular}
\caption{
 The input setup for generating the pseudodata sets
 used in the analyses demonstrated in this article.
 The theoretically estimated values and
 the reconstructed results are also given.
%
 $\dagger$: estimated for 130 GeV $\mchi$;
 $\ddagger$: different from Ref.~\cite{AMIDAS-CYGNUS2011}.
}
\label{tab:setup}
\vspace{-0.25cm}
\end{center}
\end{table}
\begin{figure}[t!]
\begin{center}
\vspace{-0.25cm}
\hspace*{-1.6cm}
\includegraphics[width=8cm]{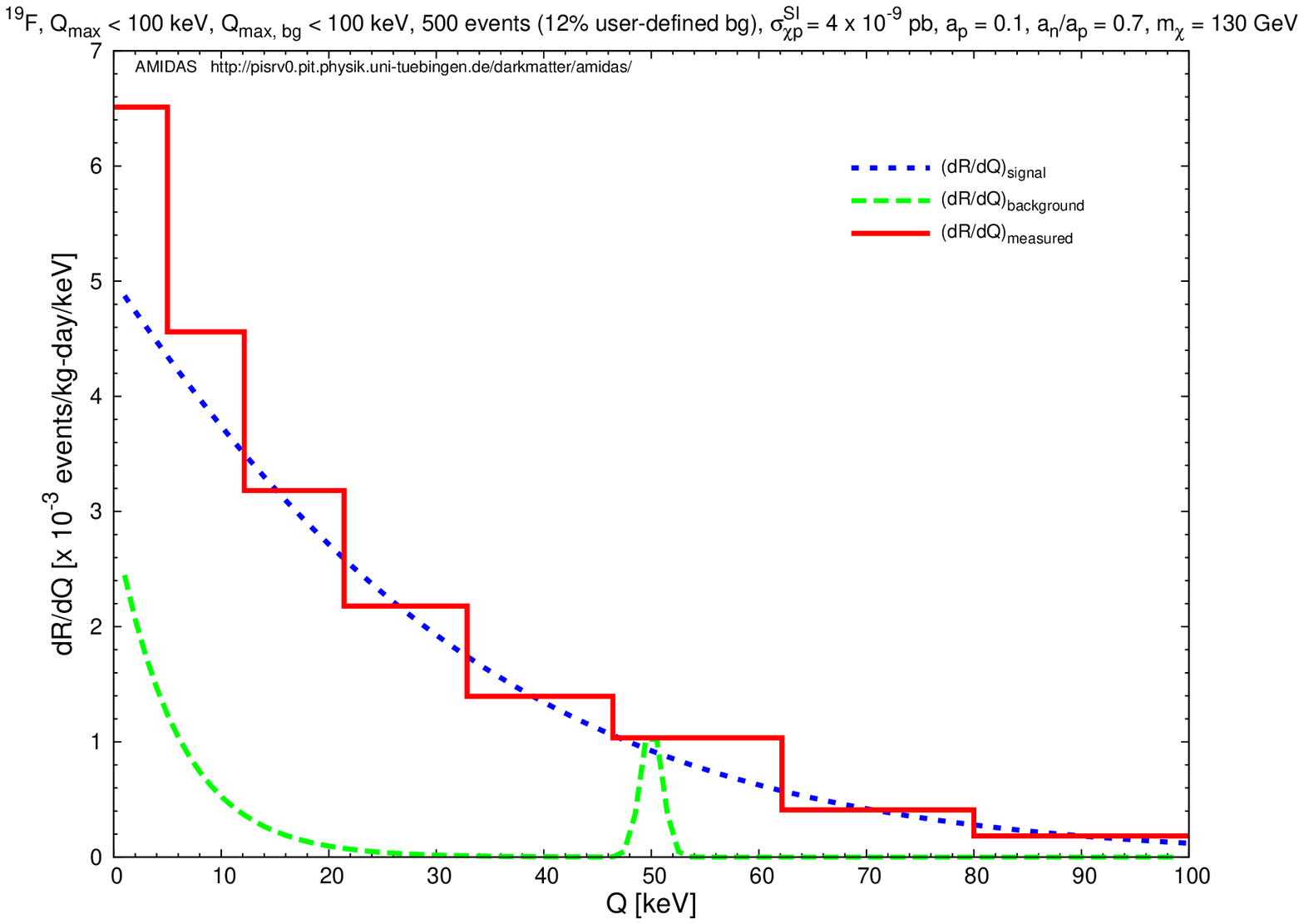}  \hspace{0.25cm}
\includegraphics[width=8cm]{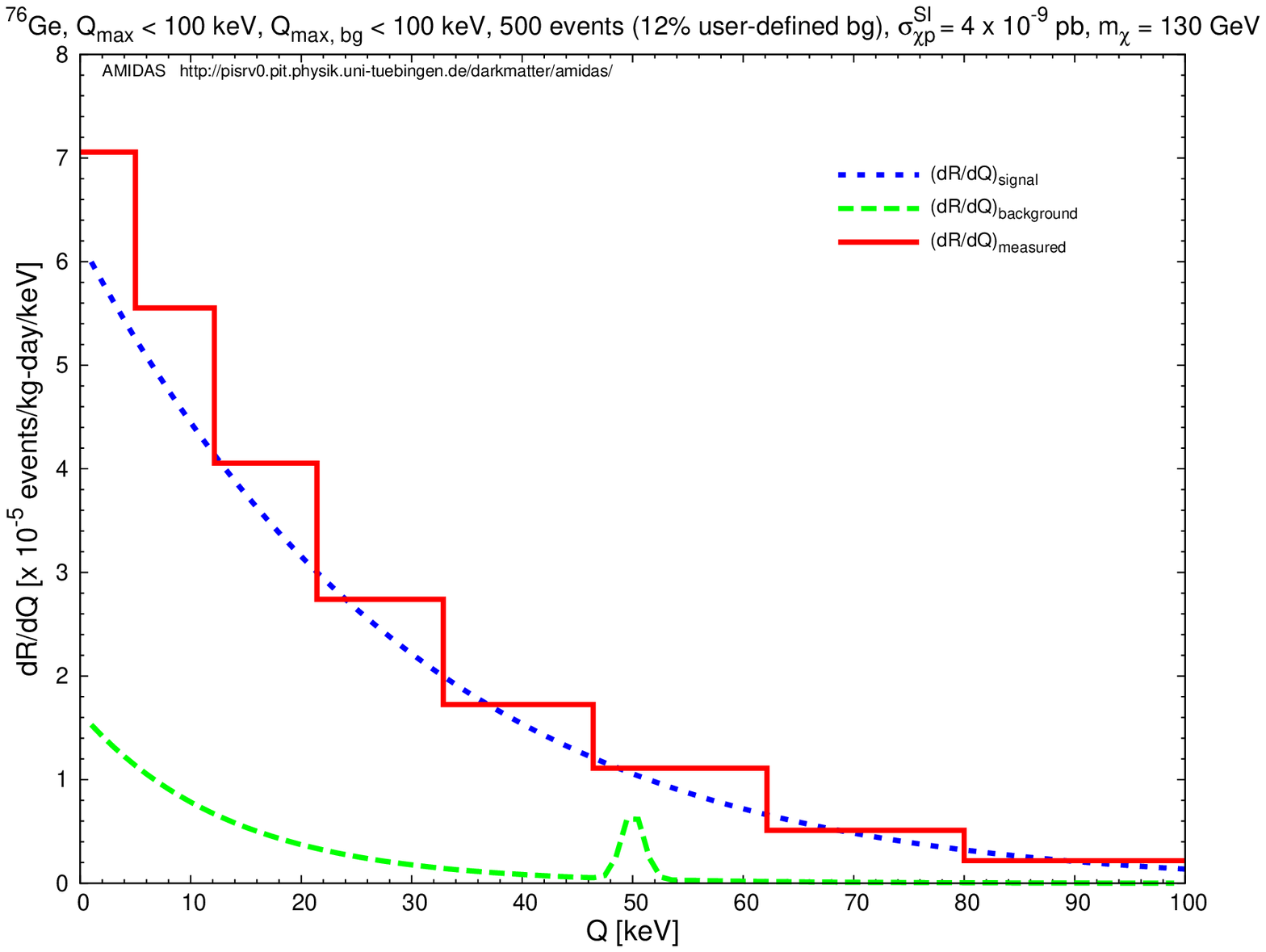} \hspace*{-1.2cm} \\
\vspace{-0.25cm}
\end{center}
\caption{
 Measured energy spectra (solid red histograms)
 for a $\rmXA{F}{19}$ (left) and a $\rmXA{Ge}{76}$ (right) targets.
 The dotted blue curves are
 the elastic WIMP--nucleus scattering spectra,
 whereas
 the dashed green curves are
 the exponential background spectra
 combined with the Gaussian excesses
 ($r_{\Gau} = 1$,
  $Q_{\rm bg, peak} = 50$ keV
  and $\sigma_{Q, {\rm bg}} = 1$ keV) and
 normalized to fit to the chosen background ratio
 $r_{\rm bg} = 12\%$.
 See the text for further details.
}
\label{fig:dRdQ-bg}
\end{figure}

 On the other hand,
 for generating background events,
 the target--dependent exponential form
 for the residue background spectrum
 introduced in Ref.~\cite{DMDDbg-mchi}:
\beq
   \aDd{R}{Q}_{\rm bg, ex}
 = \exp\abrac{-\frac{Q /{\rm keV}}{A^{0.6}}}
\label{eqn:dRdQ_bg_ex}
\eeq
 has been used.
 Here $A$ is the atomic mass number of the target nucleus;
 the power index of $A$, 0.6, is an empirical constant,
 which has been chosen so that
 the exponential background spectrum is
 somehow similar to,
 but still different from
 the expected recoil spectrum of the target nuclei
 (see Figs.~\ref{fig:dRdQ-bg})%
\footnote{
 Note that,
 among different possible choices
 the atomic mass number $A$
 has been chosen
 as the simplest, unique characteristic parameter
 in the general analytic form (\ref{eqn:dRdQ_bg_ex})
 for defining the residue background spectrum
 for {\em different} target nuclei.
 However,
 it does {\em not} mean that
 the (superposition of the real) background spectra
 would depend simply/primarily on $A$ or
 on the mass of the target nucleus, $\mN$.
}.
 Meanwhile,
 considering possible radioactivity
 with a characteristic energy,
 we combined the exponential background spectrum
 in Eq.~(\ref{eqn:dRdQ_bg_ex})
 with a Gaussian excess:
\beq
   \aDd{R}{Q}_{\rm bg, Gau}
 = \frac{r_{\Gau}}{\sqrt{2 \pi} \abrac{\sigma_{Q, {\rm bg}} / {\rm keV}}}
   \exp\bbrac{-\frac{\abrac{Q - Q_{\rm bg, peak}}^2}{2 \sigma_{Q, {\rm bg}}^2}}
\~,
\label{eqn:dRdQ_bg_Gau}
\eeq
 where $Q_{\rm bg, peak}$ and $\sigma_{Q, {\rm bg}}$
 are the characteristic energy and energy dispersion
 of this background excess,
 respectively;
 $r_{\Gau}$ is the ratio of this Gaussian excess
 to the exponential spectrum.

 In Figs.~\ref{fig:dRdQ-bg}
 I show the measured energy spectra (solid red histograms)
 for a $\rmXA{F}{19}$ (left) and a $\rmXA{Ge}{76}$ (right) targets.
 The dotted blue curves are
 the elastic WIMP--nucleus scattering spectra,
 whereas
 the dashed green curves are
 the exponential background spectra given in Eq.~(\ref{eqn:dRdQ_bg_ex})
 combined with the Gaussian excesses given in Eq.~(\ref{eqn:dRdQ_bg_Gau})
 ($r_{\Gau} = 1$,
  $Q_{\rm bg, peak} = 50$ keV
  and $\sigma_{Q, {\rm bg}} = 1$ keV
  for both targets),
 which have been normalized so that
 the ratios of the areas under the background spectra
 to those under the (dotted blue) WIMP scattering spectra
 are equal to the background--signal ratio ($r_{\rm bg} = 12\%$)
 in the whole data sets.
 The experimental threshold energies
 have been assumed to be negligible
 and the maximal cut--off energies
 are set as 100 keV.
%
 500 total events on average
 in one experiment
 have been simulated.

 It can be seen here that%
\footnote{
 More detailed illustrations and discussions
 about the effects of residue background events
 on the measured energy spectrum
 can be found in Refs.~\cite{DMDDbg-mchi, DMDDbg-ranap}.
},
 firstly,
 due to the contribution of the unrejected background events
 in low energy ranges
 ($Q~\lsim~20$ keV for $\rmXA{F}{19}$ and
  $Q~\lsim~40$ keV for $\rmXA{Ge}{76}$),
 the counting rates at the first energy bins
 (one of the two important quantities
  required in our model--independent data analyses)
 have been (strongly) overestimated.
 Secondly,
 the Gaussian background excesses around $Q = 50$ keV
 cause clearly overestimates of the event rates,
 which would not only contribute (significantly) to the estimates of
 $I_n = \sum_{\rm all~events} Q_i^{(n-1)/2} / F^2(Q_i)$ \cite{DMDDf1v},
 but also cause a larger statistical fluctuation \cite{DMDDmchi}.
 
 Nevertheless,
 our results obtained by analyzing
 pseudodata sets of $\cal O$(50) total events
 showed that,
 firstly,
 the WIMP mass given in Eq.~(\ref{eqn:mchi_rec})
 can match the input value very well:
 the deviations between the input and the reconstructed values
 with different assumed nuclear form factors
 are $\lsim$ 20\% (\mbox{$\lsim$ 30 GeV}).
 As discussed earlier,
 this indicates that,
 for the first approximation of
 giving/constraining the most plausible range of the WIMP mass
 with pretty few total events,
 the uncertainty on the nuclear form factor
 could be safely neglected.

 Secondly,
 all WIMP--nucleon couplings/cross sections
 as well as the ratios between them
 have also been reconstructed with
 $\sim$ 2\% to $\lsim$ 30\% deviations
 from the input/theoretically estimated values.
 Although the SI WIMP coupling $|f_{\rm p}|$
 estimated with the larger (input) local Dark Matter density
 (right frame of Figs.~\ref{fig:fp2})
 could be underestimated \cite{DMDDfp2},
 one can at least
 give an
 upper bound on $|f_{\rm p}|$.
 Moreover,
 by combining different methods
 for estimating different (ratios between the) WIMP couplings/cross sections,
 one could in principle observe/confirm the (in)compatibility
 between these results
 and probably correct the reconstructed values.

\section{Summary}
 In this article
 I demonstrated the data analysis results of extracted WIMP properties
 by using theoretically generated pseudodata
 for different target nuclei,
 taking into account some unrejected background events
 mixed in the analyzed data sets.
 As an extension as well as the complementarity of
 our earlier theoretical works,
 I combined reconstructed results of
 the (ratios between different) WIMP couplings/cross sections on nucleons
 to estimate each {\em individual} coupling/cross section.
 Hopefully,
 the \amidas\ package and website
 as well as this demonstration
 can help our experimental colleagues
 to analyze their {\em real} direct detection data in the near future
 and to determine (at least rough ranges of) properties of
 halo Dark Matter particles.
\subsection*{Acknowledgments}
 The author appreciates the ILIAS Project and
 the Physikalisches Institut der Universit\"at T\"ubingen
 for kindly providing the opportunity of the collaboration
 and the technical support of the \amidas\ website.
 The author would also like to thank
 the friendly hospitality of the
 Kavli Institute for Theoretical Physics China at the Chinese Academy of Sciences (KITPC)
 during the DSU workshop and the ``Dark Matter and New Physics'' program.
 This work
 was partially supported
 by the National Science Council of R.O.C.~%
 under contract no.~NSC-99-2811-M-006-031
 as well as by
 the National Center of Theoretical Sciences, R.O.C..
\section*{References}
\end{document}